\documentclass{article}
\usepackage{arxiv}
\usepackage[utf8]{inputenc} 
\usepackage[T1]{fontenc}    
\usepackage[table,xcdraw, dvipsnames]{xcolor}
\usepackage[colorlinks=true, linkcolor=blue, urlcolor=blue, citecolor=blue]{hyperref}
\usepackage{url}            
\usepackage{booktabs}       
\usepackage{amsfonts, amsmath, amssymb, bm}       
\usepackage{nicefrac}       
\usepackage{microtype}      
\usepackage{graphicx}
\usepackage{natbib}
\usepackage{doi}
\usepackage{tikz}
\usetikzlibrary{arrows.meta,positioning,calc,fit,shapes.geometric}
\usepackage{econometrics}
\usepackage{breqn}
\usepackage{enumitem}
\usepackage{subfig}
\usepackage[autostyle=false, style=english]{csquotes}
\usepackage[algo2e,ruled]{algorithm2e}
\MakeOuterQuote{"}
\usepackage{amsthm}
\usepackage[toc,page]{appendix}
\usepackage{comment}

\theoremstyle{plain}

\newtheorem{assumption}{Assumption}

\theoremstyle{definition}

\LinesNumbered
\DontPrintSemicolon 
\SetArgSty{textup}

\newcommand{\R}{\mathbb{R}}
\newcommand{\rmax}{\rho_{\max}}

\tikzset{
  block/.style={draw=black,thick,rectangle,minimum height=1.0cm,minimum width=2.6cm,
                align=center,font=\footnotesize,inner sep=4pt},
  decision/.style={draw=black,thick,diamond,aspect=2.4,align=center,font=\footnotesize,
                   inner sep=2pt},
  sum/.style={draw=black,thick,circle,inner sep=1pt,minimum size=5mm},
  line/.style={draw=black,thick,-{Latex[length=2mm]}},
}

\title{A Rollout-Based Approach to Real-Time Topology and Dispatch Control in the AI4RealNet L2RPN Benchmark}

\author{
    \href{https://orcid.org/0009-0004-0547-1941}{\includegraphics[scale=0.06]{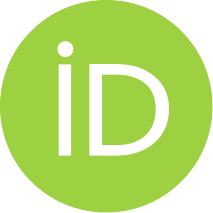}\hspace{1mm}David~Krame~Kadurha} \\
    Department of Mathematics \\
    Vrije Universiteit Amsterdam \\
	\texttt{\href{mailto:d.kramekadurha@vu.nl}{d.kramekadurha@vu.nl}}
    \And
	\href{https://orcid.org/0009-0007-4945-6120}{\includegraphics[scale=0.06]{orcid.pdf}\hspace{1mm}Robbert~van~der~Burg} \\
        Department of Mathematics \\
    Vrije Universiteit Amsterdam \\
	\texttt{\href{mailto:r.vander.burg@vu.nl}{r.vander.burg@vu.nl}} \\
    \And
	\href{https://orcid.org/0000-0001-6585-4785}{\includegraphics[scale=0.06]{orcid.pdf}\hspace{1mm}Alessandro~Zocca} \\
    Department of Mathematics \\
    Vrije Universiteit Amsterdam \\
	\texttt{\href{mailto:a.zocca@vu.nl}{a.zocca@vu.nl}}
    \And
    \href{https://orcid.org/0000-0003-1124-8821}{\includegraphics[scale=0.06]{orcid.pdf}\hspace{1mm}Sandjai~Bhulai}\\
    Department of Mathematics \\
    Vrije Universiteit Amsterdam \\
	\texttt{\href{mailto:s.bhulai@vu.nl}{s.bhulai@vu.nl}} \\
}

\renewcommand{\headeright}{ }
\renewcommand{\shorttitle}{Rollout-Based Topology and Dispatch Control}

\hypersetup{
pdftitle={Multiplicative Path Weights in Networks},
pdfauthor={Robbert van der Burg, Bernd Heidergott, Thao Le, Alessandro Zocca}
}

\begin{document}
\maketitle

\begin{abstract}
	Operating transmission networks under uncertain outages requires coordinating discrete network topology actions with continuous adjustments to generation and storage, while respecting operational constraints and real-time computational limits. We present a rollout-based controller that combines discrete and continuous decisions under the uncertain contingency policy of the L2RPN AI4RealNet 2026 challenge. Power-flow approximations and convex optimization identify the most promising actions, followed by short-horizon simulations to predict their effect on congestion, renewable curtailment, and redispatching costs. A further policy exploits the benchmark's specific attack protocol: keeping selected lines disconnected prevents some adversarial outages from being initiated. On the benchmark's $118$-bus network, the controller completed $189$ of $208$ final-test scenarios (score $64.39$) and achieved the second-best ratio of score to computation time among the five top-ranked entries, behind one that scored lower. Our results show that combining structured optimization with simulation-based action selection is effective even without a learned decision rule, and fast enough for real-time control. Code and reproduction material: \url{https://github.com/DavidKrame/RAD_agent__AI4RealNet_competition}.%
\end{abstract}


\section{Introduction}
\label{sec:intro}
Reliable operation of transmission networks requires accommodating changes in generation, demand, and line availability. Wind and solar output fluctuate with the weather, planned maintenance takes generating units and lines out of service, and sudden outages caused by extreme weather, equipment failures, or deliberate attacks disconnect lines and often require immediate action. Each change redistributes the power flow and can overload the remaining network, triggering further disconnections. The control problem is sustaining electricity supply through anticipated and sudden disruptions while limiting operating costs and curtailment of renewables.

Operators can relieve congestion by changing the network topology, redispatching generators, curtailing renewable production, or storing energy. These controls interact: a topology change redistributes power flow across the network, and a dispatch decision affects both present congestion and later decisions. The difficulty is to combine a combinatorially large set of topology actions with continuous decisions, while an overloaded network leaves little time to compute, because corrective actions must come before the disruption spreads and leads to blackouts.

Existing work offers various approaches to operating transmission networks. Security-constrained optimal power flow aims to coordinate preventive and corrective decisions across specified contingencies. Its computational challenges include contingency selection, discrete controls, and uncertainty~\citep{capitanescu2011state}. Optimal transmission switching couples generation dispatch and line switching~\citep{fisher2008switching}. These formulations establish the value of coordinated control, but their computational demands can make them difficult to solve repeatedly when rapid intervention is required. This motivates reinforcement learning (RL) approaches, which learn control policies offline that can be evaluated quickly during operation, see for instance~\citep{marot2021retrospective}. Representative approaches use hierarchical actor-critic policies~\citep{yoon2021winning} or AlphaZero-based search combined with redispatch~\citep{dorfer2022alphazero}, see also \citet{vandersar2025survey} for a wider literature review. Adversarial training has also been used to improve performance under line outages~\citep{omnes2021adversarial}.

Established methods also support control without a learned decision rule. For instance, \cite{marot2018expert} demonstrated topology-based congestion relief using an expert system. Line outage distribution factors (LODFs) are standard tools for approximating flow changes~\citep{wood2013power}, also used as inputs to learned graph-based control~\citep{degenkolb2026graph}, and rollout policies provide an established framework for comparing actions through simulated continuations~\citep{bertsekas2019reinforcement}.
Combining topology search with continuous optimization already has strong precedents. The LJN agent, derived from the winning 2023 IDF entry, combines topology management, continuous control, and operational heuristics~\citep{ljnagent}. Its continuous-control component builds on OptimCVXPY, which optimizes redispatch, curtailment, and storage under a DC approximation~\citep{optimcvxpy}. 

Building on these approaches, we focus on coordinating the controls through a common short-horizon cost, together with preventive decisions whose effects are assessed separately. We develop this controller for the L2RPN AI4RealNet 2026 challenge. The controller uses a short-horizon forward simulation to guide the selection of appropriate actions. Candidate topology and dispatch actions are generated by a combination of using topology libraries, sensitivity screening, and convex optimization. Their predicted effects are then evaluated through AC forward simulation under a common objective that accounts for congestion, renewable curtailment, and redispatch costs, following the rollout principle of \cite{bertsekas2019reinforcement}. Additional mechanisms address risks that the short horizon does not capture. Specifically, we:
\begin{enumerate}[label=(\roman*)]
    \item Specify a controller that ranks and combines topology and continuous-control candidates using predicted congestion, curtailment, and redispatch costs.

    \item Propose a convex dispatch program that uses observed AC flows as initial conditions for a DC approximation, with generator-specific ramp bounds, together with a set of rules for scheduled outage preparation and preservation of downward redispatch margins.

    \item Design and evaluate a preventive withholding policy that reduces attack exposure by keeping selected lines disconnected. Its effect depends explicitly on the benchmark rule that cancels an attack attempt when an attackable line in that region is already out of service.
\end{enumerate}

The remainder of the paper is organized as follows. Section~\ref{sec:model} defines the model and the control problem. Section~\ref{sec:architecture} builds the controller. Section~\ref{sec:validation} reports the numerical results. Section~\ref{sec:conclusion} discusses the limitations and concludes.

\section{Model}
\label{sec:model}
Let a network consist of $n$ buses $\mathcal{N}=\{1,\dots,n\}$, which can be partitioned into $Z$ disjoint control regions $\mathcal{N}=\mathcal{N}_1\cup\ldots\cup\mathcal{N}_Z$, and $m$ (possibly parallel) lines $\mathcal{L}=\{1,\dots,m\}$, where $\ell \in {\cal L}$ has endpoints $(i_\ell, j_\ell)\in {\cal N}\times {\cal N}$. We observe the network over a finite discrete-time horizon $T$, indexed by timesteps $t \in \{0, \dots, T-1\}$. Line $\ell=(i_\ell,j_\ell)$ has status $y_\ell\in\{0,1\}$, with $y_\ell=1$ when in service. It carries an active power flow $f_\ell$ and a current $I_\ell$, and its thermal limit $\bar I_\ell$ defines its loading $\rho_\ell=|I_\ell|/\bar I_\ell$, where we write $\rmax=\max_\ell\rho_\ell$. A substation with $d$ connected elements can be split across two busbars by an assignment $\beta\in\{0,1\}^{d}$, which redistributes flow among its lines without changing any continuous control (Figure~\ref{fig:substation}, Appendix~\ref{app:model}).

An action is represented by a tuple, $a=\left(\beta,y,\Delta P_g,c_r,s_u\right)$ where $\beta$ specifies the busbar assignments, $y \in \{0,1\}^m$ specifies line connections or disconnections. The tuple $(\Delta P_g,c_r,s_u)$ specifies the continuous controls: the redispatch $\Delta P_g$ of dispatchable generators $g\in\mathcal{G}\subseteq {\cal N}$, the curtailment $c_r\ge0$ of renewable units $r\in\mathcal{R}\subseteq {\cal N}$, and the power $s_u$ of storage units $u\in\mathcal{U}\subseteq {\cal N}$, positive when charging.

A generator with output $P_g^{\min} \leq P_g\leq P_g^{\max}$, and ramp limits $R_g^{\downarrow},R_g^{\uparrow}$ has one-step ramp margins
\begin{equation}
  \underline r_g=\min\big(P_g-P_g^{\min},\,R_g^{\downarrow}\big),
  \qquad
  \overline r_g=\min\big(P_g^{\max}-P_g,\,R_g^{\uparrow}\big).
  \label{eq:rampmargins}
\end{equation}
We write $a^\varnothing$ for the no-op (do-nothing action) and $a\oplus a'$ for the action that applies $a$ and then $a'$ within the same step, as the action sum of the Grid2Op platform does~\citep{grid2op}. An action is \emph{topology free} if it changes neither a busbar assignment nor a line status. A line or substation that was just acted on cannot be acted on again during a cooldown of three steps in the benchmark.

\begin{assumption}[Regional action limit]
\label{asm:admissibility}
At most one substation reconfiguration and, independently, at most one line connection or disconnection may be applied per control region per step.
\end{assumption}

We impose this limit to match the AI4RealNet challenge's admissibility rules. They permit simultaneous operations in different regions, as well as one substation reconfiguration and one line operation within the same region.

Each line has an overload counter $\kappa_\ell$, which counts the consecutive steps during which $\rho_\ell>1$ and resets to zero when $\rho_\ell \leq 1$. A line trips when $\kappa_\ell$ exceeds an allowance $\bar\kappa$. Additionally, a line is disconnected immediately when its loading exceeds a hard threshold, $\rho_\ell > \rho^{\rm hard} > 1$ (Figure~\ref{fig:protection}, Appendix~\ref{app:model}).

Unannounced outages are generated by the benchmark's regional opponent. Region $z$ has a set ${\cal B}_z$ of attackable lines, and ${\cal B}= \cup_{z=1}^Z {\cal B}_z$. At reset, each region draws a hidden schedule of attack occurrences. Idle times are geometrically distributed with mean 30 hours and durations have mean 2 hours, so that occurrences are nominally 32 hours apart. A region's opponent is called only when none of its own attacks is in progress. At such a call, an overdue occurrence is consumed. If \textit{every} line of ${\cal B}_z$ is in service, a target is proposed with probability proportional to $4^{k/(|\mathcal{B}_z|-1)}$, where $k\in\{0,\dots,|\mathcal{B}_z|-1\}$ ranks the lines from least to most loaded. If any line of ${\cal B}_z$ is out of service, the attempt is aborted and the countdown is not reset, so each subsequent call consumes a further occurrence until the schedule is empty. A proposal can still be rejected by the benchmark's duration and budget checks. An accepted attack is called \textit{realized}.

At decision step $t \in \{0, \dots, T-1\}$, the controller observes $o_t$, which contains the topology, flows, loadings, dispatch state, counters, cooldowns, a deterministic forecast of loads and renewable output over the next hour, and the steps $\tau_\ell(o_t)$ before the next scheduled outage, that cannot be prevented or canceled, of each line $\ell$. It compares candidate actions with a forward simulation model $\Gamma$ initialized from observation $o_t$. For each candidate action $a$, it simulates applying $a$ once and following the no-op policy thereafter, i.e., $o_1^a = \Gamma(o_t, a)$ and, for $h\geq 1$, $o_{h+1}^a = \Gamma(o_h^a, a^\varnothing)$, where $o_h^a$ is the \textit{simulated state} of the network $h$ steps ahead. It uses forecast injections and AC power flow, applies the overload protection, but excludes the stochastic contingency process. We write $\rho_\ell(o)$, $\kappa_\ell(o)$, and $\rmax(o)$ for the corresponding quantities in any observed or simulated state $o$.

An episode is \textit{completed} if the controller successfully survives all $T$ time steps. It terminates early if the simulator reports a backend failure, e.g., no feasible solution for the AC power flow or an infeasible redispatch, indicating an inability to satisfy generator output limits, ramp limits, and power balance simultaneously. The controller is evaluated by the score of a scenario, $0.8 \, C_{\mathrm{op}}+0.2 \, C_{\mathrm{lc}}$, with $C_{\mathrm{op}}, C_{\mathrm{lc}} \in [-100,100]$. The operation score $C_{\rm op}$ decreases with the normalized operating cost, which counts losses, flexibility, and the blackout cost after an early end; it is $-100$ for immediate failure, $0$ for the no-op, and $80$ for a policy that only reconnects lines on a network without overload protection. The low-carbon score $C_{\rm lc}$ increases with the share of available renewable energy that is used.

\section{Control Architecture}
\label{sec:architecture}

This section builds the controller from its components and assembles them into the control law in Section~\ref{sec:overview}. The controller evaluates candidate actions by AC forward simulation and proposes some of them with linear sensitivity factors. Appendix~\ref{app:notation} lists the main symbols and parameter values.

\subsection{Network sensitivity factors}
\label{sec:sensitivity}
The linear sensitivity factors rest on the DC approximation of the power flow~\citep{stott2009dc}, which neglects reactive power, voltage variation, and losses. Under this approximation and for a fixed topology, the active power flows $f$ are linear in the net injections at the buses, that is, generation minus consumption. Two standard factors follow from this linear model~\citep{wood2013power}.

The power-transfer distribution factor (PTDF) $\Phi_{\ell,i}$ is the change of the flow on line $\ell$ when one unit of power is injected at bus $i$ and withdrawn at a fixed reference bus. These factors form the matrix $\Phi\in\R^{m\times n}$. By linearity, a change of the injections that sums to zero changes the flows by $\Phi$ times that change. The dispatch program of Section~\ref{sec:dispatch} uses this relation.

The line-outage distribution factor (LODF) $L_{\ell,e}$ is the change of the flow on line $\ell$ when line $e$ is removed, per unit of the flow $f_e$ that $e$ carried. Given $e=(i_e,j_e)$,
\begin{equation}
  L_{\ell,e} \;:=\; \frac{\Phi_{\ell,i_e}-\Phi_{\ell,j_e}}{1-(\Phi_{e,i_e}-\Phi_{e,j_e})},
  \qquad
  f_\ell^{(e)} \;=\; f_\ell \;+\; L_{\ell,e}\,f_e,
  \qquad \ell\ne e,
  \label{eq:lodf}
\end{equation}
where $f_\ell^{(e)}$ is the flow on line $\ell$ after the removal. The denominator vanishes exactly when $e$ is a bridge, a line whose removal splits the network. The line-opening screen of Section~\ref{sec:candidates} and the dispatch program of Section~\ref{sec:dispatch} use these factors. Candidates proposed with either factor are then checked with the AC forward model $\Gamma$ of Section~\ref{sec:model}.

\subsection{Evaluation of an action}
\label{sec:ranking}

We evaluate each action over $H=2$ decision steps, corresponding to 10 minutes in the benchmark. The two functions below evaluate an action $a$ at $o_t$ on the simulated states $o^a_1,\dots,o^a_H$ of Section~\ref{sec:model}. Both equal $+\infty$ if a simulated step ends the episode. The first function is the peak loading,
\begin{equation}
  F_H(o_t,a) \;=\; \max_{h=1,\dots,H}\rmax(o^a_h).
  \label{eq:peak}
\end{equation}
The second function depends on a state cost. With $[\cdot]_{+}=\max(\cdot,0)$, an overload weight $\lambda\gg1$, and a counter weight $\mu>1$, the cost of a state $o$ is
\begin{equation}
  \Psi(o) \;=\; -\sum_\ell\big[1-\rho_\ell(o)\big]_+^2
  \;+\;\lambda\sum_\ell\Big(1+(\mu-1)\frac{\min(\kappa_\ell(o),\bar\kappa)}{\bar\kappa}\Big)\big[\rho_\ell(o)-1\big]_+^2.
  \label{eq:margin}
\end{equation}
The first sum rewards spare capacity, so candidates without overload are still ranked. The second penalizes overload with the large weight $\lambda$, increased up to $\mu$ times as a line's counter approaches its allowance. The cost-to-go approximation is
\begin{equation}
  J_H(o_t,a) \;=\; \max_{h=1,\dots,H} \Psi(o^a_h) \;+\; \gamma_1\, \chi(o^a_1)
    \;+\; \gamma_2\, \Omega(o^a_1)\,\mathbf{1}\{\rmax(o_t)<\rho^{\mathrm{safe}}\},
  \label{eq:ranking}
\end{equation}
where $\gamma_1,\gamma_2>0$ are weights, $\chi\in[0,1]$ is the curtailed share of renewable potential, $\Omega=\sum_g\pi_g|\Delta P_g|$ is the standing redispatch weighted by normalized generator costs $\pi_g$, and $\rho^{\mathrm{safe}}<1$ marks an uncongested network. A small tolerance $\delta>0$ defines strict improvement in $J_H$.

\subsection{Convex dispatch subproblem}
\label{sec:dispatch}

For a fixed topology, the continuous controls $\Delta P=(\Delta P_g)_{g\in\mathcal{G}}$, $c=(c_r)_{r\in\mathcal{R}}$, and $s=(s_u)_{u\in\mathcal{U}}$ enter the flows linearly through the PTDF matrix $\Phi$ of Section~\ref{sec:sensitivity}, so the joint decision is a convex quadratic program. Let $f^{\mathrm{obs}}$ be the observed AC flows, $\bar f^{\mathrm{obs}}_\ell=|f^{\mathrm{obs}}_\ell|/\rho_\ell(o_t)$ the active power that corresponds to the thermal limit at the observed operating point, and $A_g,A_c,A_s$ the maps from generators, renewable units, and storage units to the first busbar of their substation. The program is
\begin{align}
  \min_{\Delta P,\,c,\,s,\,\xi,\,\zeta} \quad & \eta_\xi\|\xi\|_2^2 + \eta_P\|\Delta P\|_2^2 + \eta_c\mathbf{1}^\top c + \eta_s\|s\|_2^2 + \eta_\zeta\mathbf{1}^\top\zeta \label{eq:qp-obj}\\
  \text{s.t.}\quad & \hat f = f^{\mathrm{obs}} + \Phi\,\Delta i, \quad \Delta i=A_g\Delta P - A_c c - A_s s, \quad \mathbf{1}^\top\Delta i=0,
    \label{eq:qp-ptdf}\\
  & -\underline r\le\Delta P\le\overline r,\quad 0\le c\le P_{\mathcal{R}},\quad \underline s\le s\le\overline s,\quad \xi,\zeta\ge0,
    \label{eq:qp-bounds}\\
  & |\hat f_\ell| \le \rho^{\mathrm{tgt}}\bar f^{\mathrm{obs}}_\ell+\xi_\ell \quad (\ell\in\mathcal{M}),
    \qquad |\hat f_\ell+L_{\ell,e}\hat f_e| \le \bar f^{\mathrm{obs}}_\ell+\zeta_{\ell e} \quad (\ell,e)\in\mathcal{Q},
    \label{eq:qp-limits}
\end{align}
with $\hat f$ the predicted flows, $\underline r,\overline r$ the ramp margins~\eqref{eq:rampmargins}, $P_{\mathcal{R}}$ the renewable outputs, $\underline s,\overline s$ the storage power limits, $\xi,\zeta$ the slacks of the two limits, $\eta_\xi,\eta_P,\eta_c,\eta_s,\eta_\zeta>0$ the weights, $\rho^{\mathrm{tgt}}<1$ the target loading, and $\mathcal{M}=\{\ell:y_\ell=1,\ \rho_\ell(o_t)>\rho^{\mathrm{tgt}}\}$ the monitored lines. The set $\mathcal{Q}$ secures scheduled outages: it pairs each line $e$ such that $0<\tau_e(o_t)\le\tau^{\mathrm{sec}}$, for the steps $\tau_e(o_t)$ before its next scheduled outage (Section~\ref{sec:model}) and a window $\tau^{\mathrm{sec}}$, with the lines $\ell$ whose post-outage flow predicted by~\eqref{eq:lodf} exceeds their limit, $|f^{\mathrm{obs}}_\ell+L_{\ell,e}f^{\mathrm{obs}}_e|>\bar f^{\mathrm{obs}}_\ell$. 
The weights satisfy $\eta_P,\eta_s\ll\eta_\zeta<\eta_c\ll\eta_\xi$. Overload relief therefore comes first, redispatch and storage are cheap, and relaxing a security limit costs less than curtailing to meet it. Constraint~\eqref{eq:qp-ptdf} anchors the flow to the observed AC flows and uses $\Phi$ only for the change the controls cause. The bounds~\eqref{eq:qp-bounds} keep each generator's request within its own ramp margin. The controller also uses OptimCVXPY~\citep{optimcvxpy}, a DC optimal power flow over redispatch, curtailment, and storage (Sections~\ref{sec:candidates} and~\ref{sec:overview}). Unlike the program above, it computes base flows from the DC approximation and decides per bus rather than per generator. 
The program is solved when $\rmax(o_t)\ge\rho^{\mathrm{tgt}}$. Its solution $a^u$, with set points clamped to their limits, is kept only if $F_H(o_t,a^u)<F_H(o_t,a^\varnothing)-\epsilon_F$, for a tolerance $\epsilon_F>0$.

\subsection{Candidate generation and composition}
\label{sec:candidates}

The candidate search runs when the forecast peak loading is high (Section~\ref{sec:overview}). It targets the congested lines $\mathcal{L}^\dagger(o_t):=\{\ell:\rho_\ell(o_t)>\rho^\dagger\}$, or the three most loaded lines if this set is empty. It builds a set $C_z(o_t)$ of topology actions for each region $z$ containing such a line, and one set $C_c(o_t)$ of continuous actions. Note that all candidates satisfy Assumption~\ref{asm:admissibility} and three sources feed $C_z$.
\begin{enumerate}[label=(\roman*)]
\item The offline library $\mathcal{T}_{\mathrm{off}}$ of single-substation bus assignments that relieved stressed states in simulation, limited to the substations of region $z$ at the ends of congested lines and to the six substations with the most connected elements.

\item A local pool $\mathcal{T}_{\mathrm{loc}}(o_t)$ of configurations with one to three elements on the second busbar and all others on the first, at the same substations and the neighbors of the most loaded line (Appendix~\ref{app:candidates}). A second run of the whole search uses this pool instead of the library, and its result is played only if its $F_H$ is at least $\epsilon_{\mathrm{loc}}$ below that of the first.

\item At most three line openings, ranked by the peak loading predicted after the opening, $q_e=\max_{\ell\ne e}|f_\ell^{(e)}|/\bar f_\ell$ with $\bar f_\ell=\sqrt{3}\,v_\ell\bar I_\ell$, where $f_\ell^{(e)}$ is given by~\eqref{eq:lodf} applied to the observed flows $f^{\mathrm{obs}}$, and $v_\ell$ is the observed line-to-line voltage of the three-phase network, so that $\bar f_\ell$ is the thermal limit in MW at unit power factor (Appendix~\ref{app:candidates}). Line $e$ is kept only if $q_e<1$. Bridges, lines under cooldown, and withheld lines are excluded.
\end{enumerate}
The continuous set $C_c$ includes ramp-limited redispatch steps, storage charging and discharging, curtailment near the most-loaded lines, OptimCVXPY when $\rmax(o_t)>1$, and the dispatch solution of Section~\ref{sec:dispatch}. With $C(o_t)=C_c(o_t)\cup\bigcup_z C_z(o_t)$, the search starts from
$
  a^\star \;=\; \operatorname*{arg\,min}_{a\in C(o_t)} \; J_H(o_t,a).
  \label{eq:argmin}
$
If $J_H(o_t,a^\star)\ge J_H(o_t,a^\varnothing)-\delta$, the search returns $a^\varnothing$, provided that $F_H(o_t,a^\varnothing)<\rho^{\mathrm{res}}$. Otherwise, it returns the first fallback whose simulated trajectory survives the horizon (Appendix~\ref{app:candidates}).

The search combines $a^\star$ with the best actions of the other regions and with the best continuous action. Let $a^\star_z$ be the best element of $C_z(o_t)$ under $J_H$, and $a^\star_c$ the best element of $C_c(o_t)$. The search visits the regions not used by $a^\star$ by increasing $J_H(o_t,a^\star_z)$ and accumulates the actions: starting from $\bar a=a^\star$, it sets $\bar a\gets\bar a\oplus a'$, first for $a'=a^\star_z$ in each region visited, then for $a'=a^\star_c$ if it is topology free and differs from $a^\star$, whenever
$
  J_H\big(o_t,\,\bar a\oplus a'\big) \;<\; J_H(o_t,\bar a) \;-\; \delta.
  \label{eq:compose}
$
The composed action $\bar a$ satisfies Assumption~\ref{asm:admissibility}: each region contributes at most one element of $C_z(o_t)$, and $a^\star_c$ is topology free.

\subsection{Preventive contingency-exposure reduction}
\label{sec:preventive}

Under the benchmark protocol, no new adversarial outage is initiated in region $z$ if any line of $\mathcal{B}_z$ is out of service. This motivates holding one line $w_z\in\mathcal{B}_z$ out of service in each region. Each $w_z$ is chosen offline as the line with the lowest mean (measured over representative states) of the simulated peak post-disconnection loading, among the lines whose removal keeps the network connected and makes no other line of $\mathcal{B}$ a bridge.

Algorithm~\ref{alg:veto} applies $\mathcal{W}=(w_1,\dots,w_Z)$ with two loading thresholds, $\rho^{\mathrm{arm}}\le\rho^{\mathrm{yld}}$, for opening the withheld lines and for holding their reconnection. The action $a^{\mathcal{W}}$ that opens the withheld lines is triggered by the loading threshold alone. It is neither ranked by $J_H$ nor checked by simulation, so it may be played even when removing a withheld line is unsafe. Table~\ref{tab:validation} measures its net effect, including this risk.

\begin{algorithm2e}[t]
\caption{Preventive withholding policy}
\label{alg:veto}
\DontPrintSemicolon\LinesNumbered
\KwIn{observation $o_t$, withheld set $\mathcal{W}$, relief action $a^{\mathrm{rel}}$ of the control law (Section~\ref{sec:overview})}
\If{$\rmax(o_t)\le\rho^{\mathrm{arm}}$ and $D=\{w_z\in\mathcal{W}: y_{w_z}=1,\ \text{no cooldown on } w_z\}\ne\varnothing$}{
  $a^{\mathcal{W}}\gets$ disconnect every line in $D$\;
  \eIf{$a^{\mathrm{rel}}$ changes no line status and $J_H(o_t,a^{\mathcal{W}}\oplus a^{\mathrm{rel}})\le J_H(o_t,a^{\mathcal{W}})+\delta$}{     \KwRet{$a^{\mathcal{W}}\oplus a^{\mathrm{rel}}$}\nllabel{ln:veto-return}\;   }{     \KwRet{$a^{\mathcal{W}}$}\;   }
}
\If{$\rmax(o_t)\le\rho^{\mathrm{yld}}$}{
    do not reconnect any line that is the only disconnected line of its set $\mathcal{B}_z$\;
}
\end{algorithm2e}

\label{sec:revalidation}
Like $\mathcal{W}$, several other parameters of the controller are set offline from the training scenarios (the loading thresholds of Table~\ref{tab:parameters}, the library $\mathcal{T}_{\mathrm{off}}$, and the offline record of preparatory actions of Section~\ref{sec:maintenance}). When new data become available, as in the challenge's retraining step, the controller tests an alternative set of these parameters against the incumbent on the same scenarios and accepts it only if the paired score gain is significant (Appendix~\ref{app:revalidation}). In the submitted controller, only $\mathcal{W}$ is retested, against one alternative $\mathcal{W}'$.

\begin{figure}[t]
\centering
\begin{tikzpicture}[every node/.style={font=\footnotesize}]
\node[block,minimum width=1.8cm,minimum height=0.62cm] (obs) {Observe\\state $o_t$};
\node[decision,aspect=3.2,inner sep=1pt,right=7mm of obs] (maint) {Scheduled outage\\imminent, severe?};
\node[block,right=9mm of maint,minimum width=3.4cm,minimum height=0.62cm] (preposition) {Apply a posture\\(Section~\ref{sec:maintenance})};
\node[decision,aspect=3.2,inner sep=1pt,below=1mm of maint] (prevent) {Withholding\\triggered?};
\node[block,right=9mm of prevent,minimum width=3.4cm,minimum height=0.62cm] (withhold) {Withhold lines\\(Algorithm~\ref{alg:veto})};
\node[block,right=9mm of withhold,minimum width=1.6cm,minimum height=0.62cm] (apply) {Apply $a_t$};
\node[block,below=1.2mm of prevent,minimum width=2.5cm,minimum height=0.62cm] (gen) {Relief by\\regime};
\node[block,right=6mm of gen,minimum width=2.5cm,minimum height=0.62cm] (rank) {Combine with\\reconnection};
\node[block,right=6mm of rank,minimum width=2.5cm,minimum height=0.62cm] (compose) {Extended-horizon\\safeguard};
\draw[line] (obs)--(maint);
\draw[line] (maint)--node[above,font=\scriptsize]{yes}(preposition);
\draw[line] (maint)--node[left,font=\scriptsize]{no}(prevent);
\draw[line] (prevent)--node[above,font=\scriptsize]{yes}(withhold);
\draw[line] (prevent)--node[left,font=\scriptsize]{no}(gen);
\draw[line] (gen)--(rank);
\draw[line] (rank)--(compose);
\draw[line] (preposition.east) -| (apply.north);
\draw[line] (withhold)--(apply);
\draw[line] (compose.east) -| (apply.south);
\end{tikzpicture}
\caption{One control cycle. The scheduled-outage look-ahead and preventive withholding can end the cycle early. Otherwise, reactive control computes an action, which the extended-horizon safeguard may replace.}
\label{fig:overview}
\end{figure}
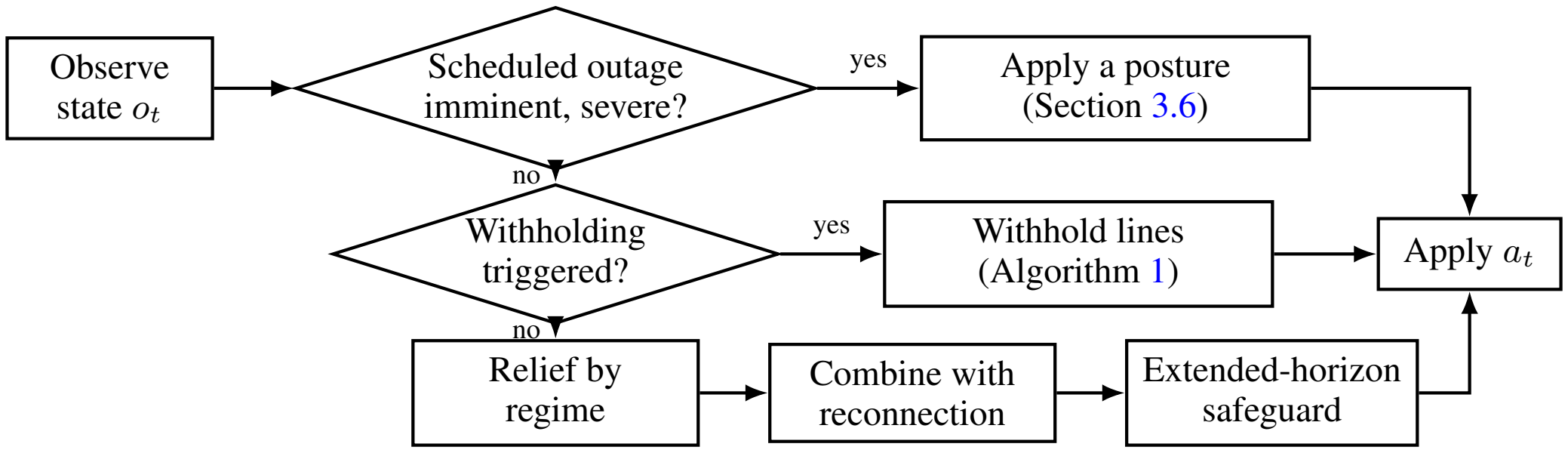

\subsection{Safeguards and control law}
\label{sec:safeguards}

\label{sec:maintenance}
The scheduled-outage look-ahead prepares the network for announced outages. Let $e$ be the in-service line with the nearest announced outage within $\tau^{\mathrm{lead}}$ steps, and let $a^{-e}$ be the action that disconnects $e$. Once per line and episode, the controller simulates $\Gamma(o_t,a^{-e})$, the outage applied now. If the simulated step does not end the episode and keeps $\rmax$ below $\rho^{\mathrm{surv}}$, nothing is done. Otherwise the controller applies a preparatory action, a \emph{posture}, taken from an offline record of postures that survived this outage when one exists, and from its own candidates otherwise (Algorithm~\ref{alg:maintenance}, Appendix~\ref{app:lookahead}).

\label{sec:flexfloor}
Margin preservation protects the downward room of the fleet, $\underline R=\sum_g\underline r_g$, with $\underline r_g$ from~\eqref{eq:rampmargins}. Returning redispatch to the economic set point can leave units at their minimum, where $\underline r_g=0$. On the $764$ local runs, $\underline R$ at the end of an episode is lower in infeasible redispatch failures than in completed episodes (medians $79$ and $151$\,MW), whereas the peak loading is not. Before that return, when $\rmax(o_t)\le\rho^{\mathrm{safe}}$, the controller shifts output from units with full downward room to units near their minimum, at zero net change (Appendix~\ref{app:lookahead}).

\label{sec:safeguard}
The extended-horizon safeguard checks the action $a_t$ of reactive control (Section~\ref{sec:overview}) over a longer horizon $H'>H$, with $S_{H'}(o_t,a)$ the number of steps, up to $H'$, that $a$ followed by no-ops survives in simulation. When $\rmax(o_t)>1$, some counter is positive, and $S_{H'}(o_t,a_t)<H'$, it ranks $a_t$ and the topology candidates of the overloaded regions by $S_{H'}$, with ties broken by lower peak loading, and plays the best one only if it survives longer than both $a_t$ and $a^\varnothing$ (Appendix~\ref{app:lookahead}).

\label{sec:overview}
Figure~\ref{fig:overview} assembles these components into one control cycle. Reactive control either applies a guarded-line rule for two lines of the credible set (Appendix~\ref{app:candidates}) or combines a relief action $a^{\mathrm{rel}}$ with a reconnection $a^{\mathrm{rec}}$.

The relief action $a^{\mathrm{rel}}$ depends on the operating regime. In the comfortable regime, $\rmax(o_t)\le\rho^{\mathrm{safe}}$, it is housekeeping. Above $\rho^{\mathrm{safe}}$, it is preventive dispatch or $a^\varnothing$ while the forecast peak loading of the no-op stays below a threshold $\rho^{\mathrm{dgr}}>1$, $F_H(o_t,a^\varnothing)<\rho^{\mathrm{dgr}}$ (the preventive regime), and the result of the candidate search (Section~\ref{sec:candidates}) otherwise.

Housekeeping lifts curtailment, sends a storage request, preserves ramp margin (Section~\ref{sec:flexfloor}), and returns redispatch toward its economic set point, each only if the forecast stays at or below $\rho^{\mathrm{safe}}$. Preventive dispatch tries OptimCVXPY on the current state, then on the state after the most likely contingency, each under a forecast check (Appendix~\ref{app:candidates}). A comfortable state goes to the candidate search if an outage is announced inside the forecast horizon and $F_H(o_t,a^\varnothing)=+\infty$. The dispatch solution of Section~\ref{sec:dispatch} replaces $a^{\mathrm{rel}}$ whenever its peak loading $F_H$ is lower by more than $\epsilon_F$. The reconnection $a^{\mathrm{rec}}$ is the recoverable line with the lowest one-step peak loading, kept if that loading is no worse than the no-op's within a small tolerance. Reactive control returns the element of $\{a^{\mathrm{rec}}\oplus a^{\mathrm{rel}},\,a^{\mathrm{rec}},\,a^{\mathrm{rel}}\}$ with the lowest $J_H$.

\section{Numerical Results}
\label{sec:validation}

We validate on the L2RPN AI4RealNet environment, built on the Grid2Op platform~\citep{grid2op} and derived from the L2RPN IDF 2023 edition of the same benchmark family~\citep{pavao2025ai}. The underlying network is a $118$-bus system with $186$ lines and $62$ generators, simulated in five-minute steps over one week, with a credible set of $22$ lines in three regions.

\paragraph{Local ablations.} We switch off one component at a time in the submitted code and run each variant on the same $764$ local scenarios and seeds (Table~\ref{tab:validation}). These runs recompute the sensitivity factors for the current topology at each step. On the challenge platform, the code recomputed them from an environment object that the platform does not step, so they remained those of the initial topology. The seeds fix the attack schedules, but which attacks are realized depends on the controller's actions (Section~\ref{sec:model}), so each row measures the total effect of a component under this attack protocol. The table also reports three reference controllers: the no-op, a policy that only reconnects lines, and OptimCVXPY alone with its published settings. The configurations and per-scenario records of all runs in Table~\ref{tab:validation}, and the script that produces it, are in the code repository.

\begin{table}[t]
\centering
\caption{Paired runs on $764$ local scenarios; each middle row removes one component. Lost and gained count the scenarios completed only with and only without the component, and $p$ is the exact two-sided McNemar test. The last two columns count failures by end cause; one more run without OptimCVXPY ended on an illegal action.}
\label{tab:validation}
\footnotesize
\begin{tabular}{@{}lrrrrrr@{}}
\toprule
controller & completed & lost & gained & $p$ & backend & infeasible \\
\midrule
full controller & $706$ & & & & $51$ & $7$ \\
\midrule
without the dispatch program (\S\ref{sec:dispatch}) & $682$ & $43$ & $19$ & $0.003$ & $68$ & $14$ \\
without OptimCVXPY, all uses (\S\ref{sec:dispatch}) & $631$ & $89$ & $14$ & $<10^{-13}$ & $115$ & $17$ \\
without secured outages $\mathcal{Q}$ (\S\ref{sec:dispatch}) & $700$ & $11$ & $5$ & $0.21$ & $56$ & $8$ \\
without the local pool (\S\ref{sec:candidates}) & $689$ & $42$ & $25$ & $0.050$ & $64$ & $11$ \\
without LODF line openings (\S\ref{sec:candidates}) & $702$ & $8$ & $4$ & $0.39$ & $52$ & $10$ \\
without opening withheld lines (\S\ref{sec:preventive}) & $679$ & $60$ & $33$ & $0.007$ & $77$ & $8$ \\
without the scheduled-outage look-ahead (\S\ref{sec:maintenance}) & $689$ & $23$ & $6$ & $0.002$ & $69$ & $6$ \\
without margin preservation (\S\ref{sec:flexfloor}) & $697$ & $34$ & $25$ & $0.30$ & $54$ & $13$ \\
without the extended-horizon safeguard (\S\ref{sec:safeguard}) & $688$ & $34$ & $16$ & $0.015$ & $65$ & $11$ \\
\midrule
reference: no-op & $152$ & & & & $612$ & $0$ \\
reference: reconnection only & $92$ & & & & $672$ & $0$ \\
reference: OptimCVXPY alone & $70$ & & & & $209$ & $485$ \\
\bottomrule
\end{tabular}
\end{table}

Six components raise the number of completed scenarios at $p<0.05$, and secured outages and LODF line openings have no detectable effect. With the dispatch program, infeasible redispatch failures fall from $14$ to $7$. OptimCVXPY alone completes only $70$ scenarios, most of them failing on infeasible redispatch, whereas inside the controller its proposals are played only after a simulated check. Switching off the opening of withheld lines, the first branch of Algorithm~\ref{alg:veto}, leaves the reconnection hold active and raises the realized attacks from $62$ to $2{,}378$, as the contingency model predicts (Section~\ref{sec:model}). These numbers measure exposure, not how well the controller withstands an attack. Reconnection alone completes fewer scenarios than the no-op and faces more realized attacks ($3{,}513$ against $1{,}226$), since reconnecting an attackable line makes its region attackable again. The look-ahead acts in only $68$ episodes, and switching it off changes the length or end cause of $47$ of them and the completion of $29$. Margin preservation is inconclusive: without it, $11$ episodes end in infeasible redispatch that do not with it, against $5$ in the other direction ($p=0.21$).

\paragraph{Challenge evaluation.} We used the challenge both to develop the controller and to test it. In its validation phase, $33$ evaluations of $14$ agent variants across $208$ unseen scenarios determined which mechanisms were retained. The retraining step applies the revalidation rule~\eqref{eq:decisionrule} to 10 scenarios from the training data provided by the challenge. During validation, the alternative withheld set $\mathcal{W}'$ led the incumbent by $16.5$ points on these scenarios and was accepted, and the validation score rose by $0.78$, within noise (paired $t=0.58$). In the final test phase, $\mathcal{W}'$ trailed by $4.2$ points, and the incumbent was kept. The submitted controller completed $184$ of $208$ validation episodes (score $72.93$) and $189$ of $208$ on the sequestered final test set (score $64.39$). The final phase had more completions but a lower normalized score; we have not yet isolated the cause. On the evaluation leaderboard, the agent had the second-best ratio of score to computation time among the five top-ranked entries ($72.93$ in $17{,}384$\,s of total agent time, against $6{,}744$ to $57{,}763$\,s for the other four). In both phases, $99\%$ of decisions took less than $0.77$\,s, and the slowest $10.2$\,s, for five-minute steps.

\label{sec:learned}
\paragraph{Learned alternatives.} Earlier versions of the controller tested learned components at three points: a proposer that adds candidate actions, a value function that ranks them, and a direct policy that selects actions without the rollout comparison. The proposer improved the score by less than the evaluation set could detect, the best value function only tied the hand-designed ranking, and the direct policy performed worse after training than before. We trained value functions to predict survival after contingencies 40–200 steps ahead, well beyond $H$, and linear and gradient-boosted classifiers fitted to these survival labels ranked the candidates of a state with an area under the ROC curve of only $0.57$ to $0.61$, against $0.5$ for a random ranking. These results concern the tested models, features, and labels, not learned control in general.

\section{Discussion and Conclusion}
\label{sec:conclusion}
This study shows that structured optimization and short-horizon simulation can coordinate topology and dispatch control. On the 2026 AI4RealNet challenge, the controller completed 189 of 208 held-out test scenarios. In paired ablations, six components each increased completions, among them the opening of withheld lines and the scheduled-outage look-ahead. The effect of margin preservation was inconclusive. These results show the value of anticipating disruptions and keeping control options for later, alongside immediate congestion relief.

The controller searches a restricted set of actions, so it can miss better ones, and its forward simulations ignore unannounced outages (attacks). Preventive withholding reduces attacks because of a benchmark rule: an attack attempt in a region is canceled when one of its attackable lines is already disconnected. An attacker that can target another line could instead exploit the redundancy lost by withholding. We tested a single network under a single attack protocol, so other networks and comparisons with similar controllers are needed to show whether the results transfer.

A natural next step is to replace the fixed attack protocol with adversaries that have explicit objectives, information, and resources, such as an attacker who adapts its timing and targets to maximize supply interruptions or operating costs. Attacker--defender models of grid protection~\citep{YUAN201483}, extended over repeated control steps, provide a starting point.

Reinforcement learning is a complementary extension to consider, although the learned components of Section~\ref{sec:learned} showed no statistically detectable benefit. A learned proposer could explore beyond the topology libraries, with optimization determining continuous controls and simulation evaluating proposals. Training against several adversaries would test whether learning improves adaptation, measured by completion, operating cost, and decision latency under matched conditions. The broader opportunity is to expand the actions the controller can discover while retaining explicit evaluation of their operational consequences.

\section*{Generative AI Use}
During the preparation of this manuscript, the authors used Claude (Opus 5.5, Anthropic) and ChatGPT (GPT-6.1 Sol, OpenAI) to assist with coding and experimental implementation, literature search, and drafting and editing parts of the manuscript. The authors reviewed and verified all AI-assisted content and take full responsibility for the content of the paper.

\section*{Acknowledgments}
D. Krame Kadurha, R. van der Burg and A. Zocca are supported by the NWO VI.Vidi.233.247 grant \textit{Power Network Optimization in the Age of Climate Extremes}~\url{https://doi.org/10.61686/GOOEL09973}.

\bibliographystyle{unsrtnat}        
\bibliography{references}           




\newpage
\begin{appendices}
\section{Network Model Illustrations}
\label{app:model}

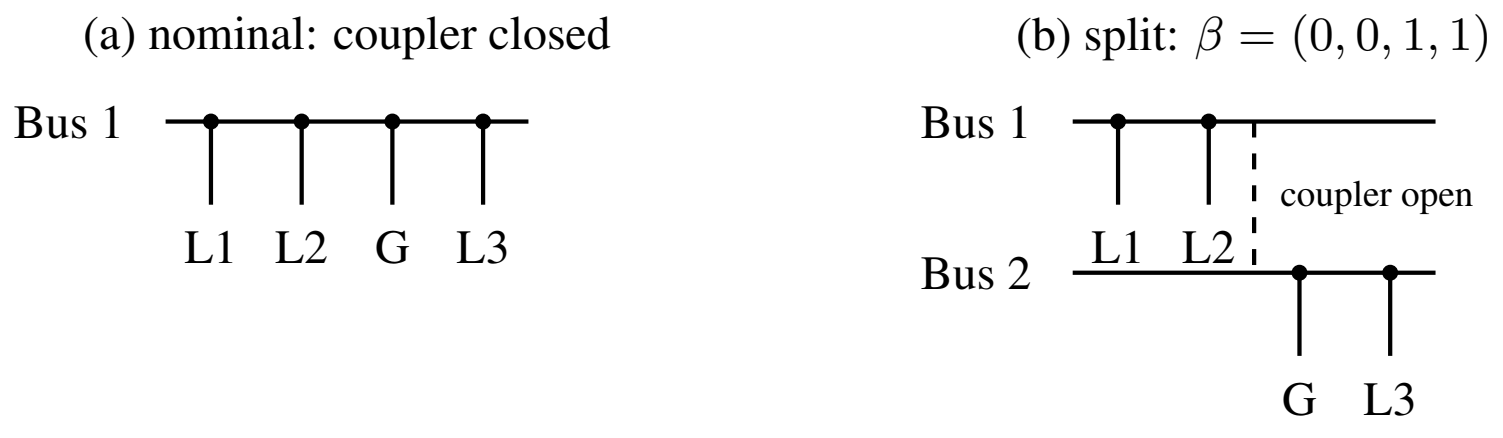
\begin{figure}[ht]
\centering
\begin{tikzpicture}[every node/.style={font=\footnotesize}]
 \begin{scope}
  \draw[thick] (0,0) -- (2.4,0);
  \node[anchor=east] at (-0.15,0) {Bus 1};
  \foreach \x/\lab in {0.3/L1,0.9/L2,1.5/G,2.1/L3}{
    \draw[thick] (\x,0) -- (\x,-0.55);
    \filldraw (\x,0) circle (1.3pt);
    \node[below] at (\x,-0.6) {\lab};
  }
  \node at (1.2,0.55) {(a) nominal: coupler closed};
 \end{scope}
 \begin{scope}[xshift=6.0cm]
  \draw[thick] (0,0) -- (2.4,0);
  \draw[thick] (0,-1.0) -- (2.4,-1.0);
  \node[anchor=east] at (-0.15,0) {Bus 1};
  \node[anchor=east] at (-0.15,-1.0) {Bus 2};
  \draw[thick,dashed] (1.2,0) -- (1.2,-1.0);
  \node[right,font=\scriptsize] at (1.25,-0.5) {coupler open};
  \draw[thick] (0.3,0) -- (0.3,-0.55); \filldraw (0.3,0) circle (1.3pt);
  \node[below] at (0.3,-0.6) {L1};
  \draw[thick] (0.9,0) -- (0.9,-0.55); \filldraw (0.9,0) circle (1.3pt);
  \node[below] at (0.9,-0.6) {L2};
  \draw[thick] (1.5,-1.0) -- (1.5,-1.55); \filldraw (1.5,-1.0) circle (1.3pt);
  \node[below] at (1.5,-1.6) {G};
  \draw[thick] (2.1,-1.0) -- (2.1,-1.55); \filldraw (2.1,-1.0) circle (1.3pt);
  \node[below] at (2.1,-1.6) {L3};
  \node at (1.2,0.55) {(b) split: $\beta=(0,0,1,1)$};
 \end{scope}
\end{tikzpicture}
\caption{Node-breaker model of a four-element substation. (a) Nominal configuration: all elements share one electrical node. (b) A topology action opens the bus coupler and reassigns generator G and line L3 to the second bus, changing the network's effective graph without any redispatch, curtailment, or storage action.}
\label{fig:substation}
\end{figure}

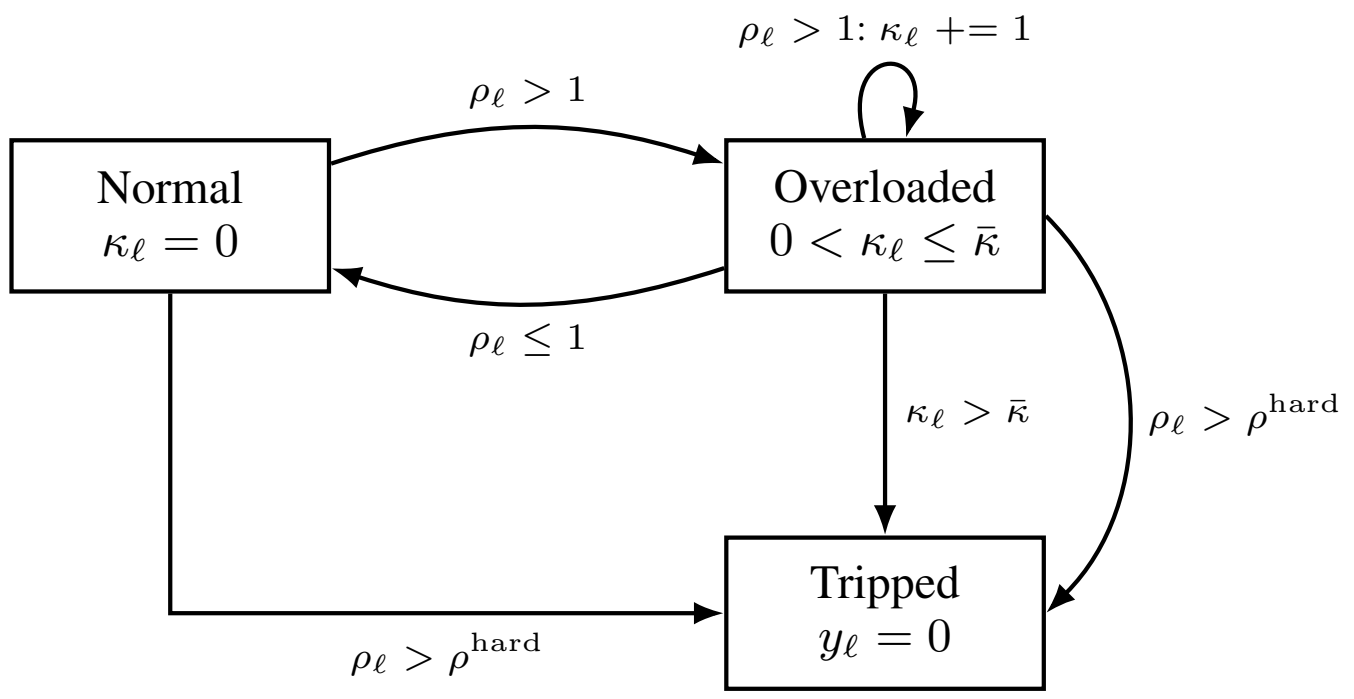
\begin{figure}[ht]
\centering
\begin{tikzpicture}[every node/.style={font=\footnotesize}]
\node[block,minimum width=2.1cm] (normal) {Normal\\$\kappa_\ell=0$};
\node[block,minimum width=2.1cm,right=26mm of normal] (counting) {Overloaded\\$0<\kappa_\ell\le \bar\kappa$};
\node[block,minimum width=2.1cm,below=16mm of counting] (tripped) {Tripped\\$y_\ell=0$};
\draw[line] (normal) to[bend left=18] node[above,font=\scriptsize]{$\rho_\ell>1$} (counting);
\draw[line] (counting) to[bend left=18] node[below,font=\scriptsize]{$\rho_\ell\le1$} (normal);
\draw[line] (counting) to[loop above,looseness=6] node[above,font=\scriptsize]{$\rho_\ell>1$: $\kappa_\ell\mathrel{+}=1$} (counting);
\draw[line] (counting) -- node[right,font=\scriptsize]{$\kappa_\ell>\bar\kappa$} (tripped);
\draw[line] (normal.south) |- node[below,font=\scriptsize,pos=0.75]{$\rho_\ell>\rho^{\mathrm{hard}}$} (tripped.west);
\draw[line] (counting.east) to[bend left=45] node[right,font=\scriptsize,align=left]{$\rho_\ell>\rho^{\mathrm{hard}}$} (tripped.east);
\end{tikzpicture}
\caption{Thermal overload protection as a state diagram. A sustained overload is tolerated for up to $\bar\kappa$ consecutive steps before a trip. An overload above the hard threshold $\rho^{\mathrm{hard}}$ trips the line immediately from either state, regardless of $\kappa_\ell$.}
\label{fig:protection}
\end{figure}

Figures~\ref{fig:substation} and~\ref{fig:protection} illustrate two parts of the model of Section~\ref{sec:model}: the substation model, and the line protection as a state diagram with three states: normal, overloaded, and tripped.

\section{Notation and Parameter Values}
\label{app:notation}

$_{}$\\
Table~\ref{tab:symbols} lists the main symbols of the paper, and Table~\ref{tab:parameters} the parameter values of the submitted controller and of the benchmark.

\begin{table}[ht]
\centering
\caption{Symbols.}
\label{tab:symbols}
\small
\begin{tabular}{@{}ll@{}}
\toprule
symbol & meaning \\
\midrule
$\mathcal{N},\mathcal{L},n,m$ & buses, lines, and their numbers \\
$\mathcal{N}_z,\mathcal{B}_z,Z$ & buses and attackable lines of region $z$, number of regions \\
$\mathcal{G},\mathcal{R},\mathcal{U}$ & dispatchable generators, renewable units, storage units \\
$y_\ell,f_\ell,I_\ell,\bar I_\ell,\rho_\ell,\rmax$ & status, active flow, current, thermal limit, loading, largest loading \\
$\kappa_\ell,\bar\kappa,\rho^{\mathrm{hard}}$ & overflow counter, its allowance, immediate trip threshold \\
$\tau_\ell$ & steps before the next scheduled outage of line $\ell$ \\
$\Phi,L$ & PTDF, LODF \\
$\Delta P_g,c_r,s_u$ & redispatch, curtailment, storage power (positive when charging) \\
$P_g,\underline r_g,\overline r_g,R_g^{\downarrow},R_g^{\uparrow}$ & generator output, downward and upward ramp margins, ramp limits \\
$o_t,a_t,a^\varnothing,\oplus$ & observation, applied action, no-op, application in the same step \\
$\Gamma,o^a_h$ & forward model, simulated state $h$ steps after $a$ \\
$F_H,J_H,\Psi,\chi,\Omega$ & peak loading, cost-to-go, state cost, curtailed share, standing redispatch \\
$S_{H'}$ & simulated survival over the horizon $H'$ \\
$C_z,C_c,C,a^\star,\bar a$ & regional and continuous candidate sets, their union, best single, composed action \\
$a^{\mathrm{rel}},a^{\mathrm{rec}},a^u,a^{\mathcal{W}},a^{-e}$ & relief, reconnection, dispatch solution, withholding, scheduled disconnection \\
$q_e,\bar f_\ell$ & screen score of line $e$, thermal limit converted to MW at unit power factor \\
$\bar f^{\mathrm{obs}}_\ell$ & thermal limit converted to MW at the observed power factor \\
$\mathcal{M},\mathcal{Q},\xi,\zeta$ & monitored lines, secured outage pairs, and their slacks \\
$w_z,\mathcal{W},\underline R$ & withheld line of region $z$, withheld set, fleet downward capacity \\
\bottomrule
\end{tabular}
\end{table}

\begin{table}[ht]
\centering
\caption{Parameter values.}
\label{tab:parameters}
\small
\begin{tabular}{@{}lll@{\hspace{2em}}lll@{}}
\toprule
symbol & value & use & symbol & value & use \\
\midrule
$H$ & $2$ & evaluation horizon & $\rho^{\mathrm{safe}}$ & $0.90$ & comfortable regime \\
$H'$ & $8$ & safeguard horizon & $\rho^{\mathrm{dgr}}$ & $1.30$ & candidate search \\
$\lambda$ & $200$ & overload weight in $\Psi$ & $\rho^{\mathrm{res}}$ & $1.20$ & last resort \\
$\mu$ & $4$ & counter weight in $\Psi$ & $\rho^\dagger$ & $0.99$ & congested lines \\
$\gamma_1$ & $0.05$ & curtailment in $J_H$ & $\rho^{\mathrm{tgt}}$ & $0.95$ & dispatch program \\
$\gamma_2$ & $0.1$ & redispatch cost in $J_H$ & $\rho^{\mathrm{arm}}$ & $0.95$ & open withheld lines \\
$\delta$ & $10^{-6}$ & strict improvement in $J_H$ & $\rho^{\mathrm{yld}}$ & $1.00$ & hold reconnections \\
$\epsilon_F$ & $10^{-4}$ & dispatch acceptance & $\rho^{\mathrm{surv}}$ & $1.30$ & outage severity \\
$\epsilon_{\mathrm{off}}$ & $0.01$ & library screening & $\tau^{\mathrm{lead}}$ & $12$ & look-ahead window \\
$\epsilon_{\mathrm{loc}}$ & $0.07$ & local pool acceptance & $\tau^{\mathrm{sec}}$ & $48$ & secured outages \\
$\eta_\xi,\eta_c,\eta_\zeta$ & $1000,50,5$ & program weights & $\Delta^{\mathrm{flex}}$ & $5$\,MW & margin preservation \\
$\eta_P,\eta_s$ & $0.03,0.1$ & program weights & $\alpha,M,N$ & $0.05,1,10$ & revalidation \\
\midrule
$\bar\kappa,\rho^{\mathrm{hard}}$ & $3,2$ & protection & $Z,|\mathcal{B}_z|$ & $3,\ (6,10,6)$ & regions \\
cooldown & $3$ steps & line and substation & $T$ & one week & five-minute steps \\
\bottomrule
\end{tabular}
\end{table}

The withheld set of the submitted controller is $\mathcal{W}=(106,131,33)$, and the alternative tested by the rule~\eqref{eq:decisionrule} is $\mathcal{W}'=(162,110,61)$. The library holds $907$ assignments at $40$ substations. The offline record of Algorithm~\ref{alg:maintenance} holds $246$ assignments of substation $79$ for the outage of line $175$. The set $\mathcal{Q}$ is limited to its $40$ most violated pairs.

\section{Reactive Control and Candidate Generation Details}
\label{app:candidates}

This appendix completes Sections~\ref{sec:overview} and~\ref{sec:candidates}. Every candidate is filtered by Assumption~\ref{asm:admissibility} and by cooldowns before being evaluated.

\paragraph{Guarded-line rule.} For two lines of the credible set whose removal was found to be especially dangerous, a library holds one bus assignment validated offline. The rule arms that posture when the line is in service, its loading lies in $[0.92,1]$, $\rmax\le1$, it is the most loaded in-service line of its region's $\mathcal{B}_z$, and the forward model shows no termination and a peak loading of at most $1.55$. The controller then plays $a^\varnothing$ while the posture is held. It restores the earlier assignment once the line's loading stays below $0.88$ for two steps, or after eight steps if it is below $0.88$ then, and it yields to reactive control if $\rmax>1.7$.

\paragraph{Housekeeping.} In the comfortable regime (Section~\ref{sec:overview}), the controller first looks for curtailment worth lifting, when more than $2\%$ of renewable potential is curtailed, and takes the largest lift that the forecast allows and that keeps the added renewable output at least $30$\,MW below the fleet's downward capacity $\underline R$ of Section~\ref{sec:flexfloor}. If it finds one, it adds one ramp-limited step of redispatch toward the economic set point if the forecast allows it, and returns. Otherwise it takes as a base action a request, for each storage unit that is not full, of a power equal to its absorption limit, sent with the sign that Grid2Op executes as discharging, if renewable output is at least $60\%$ of capacity on average, or else $a^\varnothing$. It then adds margin preservation (Section~\ref{sec:flexfloor}) if available, or else the step of redispatch.

\paragraph{Preventive dispatch.} In the preventive regime (Section~\ref{sec:overview}), the controller tries three actions in order. (i) If $\rmax>1$, the OptimCVXPY solution, played if it lowers $J_H$ below $J_H(o_t,a^\varnothing)-\delta$. (ii) An anticipatory action: if removing the most loaded in-service line of $\mathcal{B}$, the most likely target, gives a one-step peak loading above $1.7$ without termination, the same program is solved on that simulated state, and the solution is played if the one-step forecast from $o_t$ shows no termination and a peak loading of at most $\rho^{\mathrm{safe}}+0.2$. After this action, the same path stays inactive for the next eight steps outside the comfortable regime. (iii) If $\rmax>1$ and a counter is positive, the same program is solved again and played if the one-step peak loading falls below $\rmax(o_t)$. If none applies, the action is $a^\varnothing$.

\paragraph{Offline library of bus assignments.} The library $\mathcal{T}_{\mathrm{off}}$ of single-substation bus assignments is built offline. (i) For every substation with at least four connected elements, the assignments are enumerated, or randomly sampled for substations with too many assignments to enumerate. (ii) Each assignment is evaluated by one-step forward simulation on a set of stressed states, with $\rmax>0.85$. (iii) An assignment is retained only if, on at least one such state, it causes no termination and lowers the peak loading by more than $\epsilon_{\mathrm{off}}$ against the no-op. At most $60$ assignments per substation are kept, ranked by the number of states they help.

\paragraph{State-dependent local pool.} For each region $z$ and observation $o_t$, the pool $\mathcal{T}_{\mathrm{loc}}(o_t)$ of Section~\ref{sec:candidates} is built at decision time in three steps. (i) The substations $V$ are those of region $z$ at the ends of the congested lines $\mathcal{L}^\dagger(o_t)$, the hubs of region $z$ (among the six substations of the network with the most connected elements), and the neighbors in region $z$ of the endpoints of the most loaded line. (ii) Substations under cooldown are removed from $V$. (iii) The pool holds every configuration of a substation of $V$ with one or two elements on the second busbar and all others on the first. At the endpoint of the most loaded line with more elements, if it has at most ten, it also holds the configurations with three elements on the second busbar.

\paragraph{LODF-screened voluntary disconnections.} The lines kept by the screen of Section~\ref{sec:candidates} are taken in ascending order of $q_e$, at most three, and added to the pool of the first congested region.

\paragraph{Continuous templates and last resort.} The templates of $C_c(o_t)$ are a step at the ramp limit of each generator in each direction, the maximum charge and discharge of each storage unit and of all units together, filtered by state of charge, and curtailment at four levels on the renewable units near the most loaded lines. The last resort tries the redispatch and storage templates, then local curtailment, then curtailment of all renewable units, and returns the first tier with a candidate that survives the horizon.

\paragraph{Limit used by the line-opening screen.} Like $\bar f^{\mathrm{obs}}_\ell$ of Section~\ref{sec:dispatch}, $\bar f_\ell$ of Section~\ref{sec:candidates} expresses the thermal limit $\bar I_\ell$ in MW, with $v_\ell$ in kV and $\bar I_\ell$ in kA, but at unit power factor instead of the observed one. The screen needs this form because it evaluates every line after an opening, including lines that carry no flow now, for which $|f^{\mathrm{obs}}_\ell|/\rho_\ell$ is $0/0$, whereas $\sqrt3\,v_\ell\bar I_\ell$ is defined for every line.

\section{Revalidation of Fitted Parameters}
\label{app:revalidation}

The parameters set offline in Section~\ref{sec:preventive}, $\mathcal{W}$ included, are denoted by $\psi$. The parameters that depend most on the data may not suit scenarios that follow a completely new distribution, so when new data become available, as in the challenge's retraining step, the controller tests an alternative $\psi'$ against the incumbent $\psi$ on the same $N$ scenarios. Let $\Delta_i$ be the score of $\psi'$ minus the score of $\psi$ on scenario $i$, with mean $\bar\Delta$ and sample standard deviation $\hat\sigma_\Delta$. The alternative is accepted only if
\begin{equation}
  \bar\Delta \;>\; z_{\alpha/(2M)}\,\hat\sigma_\Delta/\sqrt{N},
  \label{eq:decisionrule}
\end{equation}
where $M$ is the number of alternatives, $\alpha$ the family-wise level, and $z_{\alpha/(2M)}$ the upper $\alpha/(2M)$ quantile of the standard normal distribution, a normal approximation to the paired test. At most one change is accepted, and any error or inconclusive result keeps the incumbent. The test adapts to the time available: $N$ is the number of scenarios on which the incumbent and the $M$ alternatives can be scored within the time reserved for the test. In the submitted controller, $\psi$ contains only the withheld set $\mathcal{W}$, with one alternative $\mathcal{W}'$ (Appendix~\ref{app:notation}), and one hour reserved for the test gave $N=10$. Section~\ref{sec:validation} reports both applications of this rule in the challenge, but the rule extends to all parameters in $\psi$, given enough computation time for training.

\section{Safeguard Details}
\label{app:lookahead}

\paragraph{Scheduled-outage look-ahead.} Algorithm~\ref{alg:maintenance} states the look-ahead of Section~\ref{sec:maintenance}. It runs before the other stages of the control law, once for each line whose announced outage enters the window $\tau^{\mathrm{lead}}$.

\begin{algorithm2e}[ht]
\caption{Scheduled-outage look-ahead}
\label{alg:maintenance}
\DontPrintSemicolon\LinesNumbered
\KwIn{observation $o_t$, line $e$ with $0<\tau_e(o_t)\le\tau^{\mathrm{lead}}$ not yet evaluated, severity threshold $\rho^{\mathrm{surv}}$}
$o'\gets\Gamma(o_t,a^{-e})$\;
\If{$o'$ does not end the episode and $\rmax(o')<\rho^{\mathrm{surv}}$}{
  \KwRet{no posture}\;
}
\eIf{an offline record $\mathcal{P}_e$ of postures that survived the outage of $e$ exists}{
  $A\gets\{a\in\mathcal{P}_e: J_H(o_t,a\oplus a^{-e})<+\infty\}$\;
  \If{$A\ne\varnothing$}{
    \KwRet{the highest-ranked posture in $A$}\;
  }
}{
  $C\gets$ the topology candidates of the regions of the endpoints of $e$, and OptimCVXPY\;
  $a\gets\arg\min_{a'\in C}J_H(o_t,a'\oplus a^{-e})$\;
  \If{$J_H(o_t,a\oplus a^{-e})<J_H(o_t,a^{-e})-\delta$}{
    \KwRet{$a$}\;
  }
}
\KwRet{no posture}\;
\end{algorithm2e}

The record is built offline by playing substation configurations through each scheduled outage in historical scenarios and keeping those observed to survive it. The controller prefers the record to the run-time forecast. For this class of event, ranking postures by their forecast post-outage loading favored configurations that later failed, because whether an overload clears depends on the steps after the outage, which a one-step forecast does not see. The forecast-ranked fallback is weaker, but it lets the mechanism act on outages without a record. The mechanism applies only to contingencies whose time is known in advance, since the horizon $H$ of the forecast used elsewhere is too short to value a posture against a later disconnection.

\paragraph{Margin preservation.} Units less than one ramp step above their minimum raise their output, each by up to $t_g=\min(R_g^{\downarrow}-\underline r_g,\,R_g^{\uparrow})$, and recover room. Units with $\underline r_g=R_g^{\downarrow}$ lower their output by the same total. The shift is played if $\sum_g t_g\ge\Delta^{\mathrm{flex}}$ and $F_H\le\rho^{\mathrm{safe}}$. Let $\mathcal{G}^{\mathrm{sat}}=\{g:\underline r_g=R_g^{\downarrow}\}$ be the units that give output, and let the transferred power be
\begin{equation*}
  P^{\mathrm{tr}}=\min\Big(\sum_g t_g,\ \sum_{g\in\mathcal{G}^{\mathrm{sat}}}R_g^{\downarrow}\Big).
\end{equation*}
Each unit's redispatch is its share of $P^{\mathrm{tr}}$:
\begin{equation*}
  \Delta P_g=\frac{t_g\,P^{\mathrm{tr}}}{\sum_{g'}t_{g'}}\ \ \text{if } t_g>0,
  \qquad
  \Delta P_g=-\frac{R_g^{\downarrow}\,P^{\mathrm{tr}}}{\sum_{g'\in\mathcal{G}^{\mathrm{sat}}}R_{g'}^{\downarrow}}\ \ \text{if } g\in\mathcal{G}^{\mathrm{sat}}.
\end{equation*}
The increases and the decreases sum to zero, and each redispatch respects the ramp limit of its unit.

\paragraph{Extended-horizon safeguard.} The horizon-$H$ evaluation of Section~\ref{sec:ranking} can miss a line held at mild overload, $\rho_\ell\in(1,\rho^{\mathrm{hard}})$, for several steps without tripping. Within $H$ steps, a slow approach to the allowance $\bar\kappa$ can look no worse than a short excursion. The safeguard of Section~\ref{sec:safeguard} takes as wider candidate set the action $a_t$ of reactive control and the topology candidates of the regions of the overloaded lines. It keeps the $20$ with the lowest one-step peak loading and ranks them by $S_{H'}$, with ties broken by the lower peak loading over the horizon $H'$.

\section{Computational Complexity and Deployment}
\label{app:complexity}

One control cycle has three parts. It builds the candidate set $C(o_t)$, whose size $|C(o_t)|=O(|\mathcal{T}_{\mathrm{off}}|+|\mathcal{T}_{\mathrm{loc}}(o_t)|+|\mathcal{G}|+|\mathcal{U}|)$ grows with the number of controllable units through the continuous templates but does not depend on the number of buses once the library and the locality rules of Section~\ref{sec:candidates} are fixed. It simulates each candidate over the horizon $H$, with one power-flow solve per step. It solves the quadratic program~\eqref{eq:qp-obj}--\eqref{eq:qp-limits} a bounded number of times. A sparse Newton--Raphson AC power flow on $n$ buses costs $O(n^{1.5})$ per iteration with a sparse factorization and needs few iterations near a feasible operating point. The quadratic program has $O(|\mathcal{G}|+|\mathcal{R}|+|\mathcal{U}|+|\mathcal{M}|+|\mathcal{Q}|)$ variables and constraints. It is convex, so an interior-point method solves it to any given accuracy in polynomial time. The dominant cost of one cycle is therefore
\begin{equation}
  O\big(|C(o_t)|\cdot H \cdot n^{1.5}\big),
\end{equation}
linear in the size of the candidate set and in the horizon, and polynomial in the number of buses through the power flow alone. In the challenge, the controller used $17{,}384$\,s of agent time for $208$ one-week scenarios, about $84$\,s per scenario (Section~\ref{sec:validation}).

Every decision can be reproduced offline from the same inputs, because each action is either produced by a deterministic rule or selected by comparing simulated trajectories. Deployment requires a state estimator that supplies $o_t$ at each step and a contingency list that supplies the credible set $\mathcal{B}=\mathcal{B}_1\cup\ldots\cup\mathcal{B}_Z$, as in security-constrained optimal power flow~\citep{capitanescu2011state}. It also requires historical or simulated network states, from which the offline library of bus assignments of Section~\ref{sec:candidates} and the posture record of Algorithm~\ref{alg:maintenance} are built offline. Only these offline libraries depend on the network, and the same procedures rebuild them for another network.
\end{appendices}

\end{document}